\documentclass[aps,prl,preprint,groupedaddress]{revtex4-2}
\usepackage{latexsym}
\usepackage{graphicx}
\usepackage{float}
\usepackage{xcolor,colortbl}

\begin{document}
\pagestyle{myheadings}

\title{The observation of vibrating pear shapes in radon nuclei: update}

\author{P.A.~Butler$^1$}
\email{peter.butler@liverpool.ac.uk}

\author{L.P.~Gaffney$^{1,2}$}

\author{P.~Spagnoletti$^3$}

\author{J.~Konki$^2$}

\author{M.~Scheck$^3$}

\author{J.F.~Smith$^3$}

\author{K.~Abrahams$^4$}

\author{M.~Bowry$^{3,5}$}

\author{J~.Cederk{\"a}ll$^6$}

\author{T.~Chupp$^7$}

\author{G.~de Angelis$^8$}

\author{H.~De~Witte$^9$}

\author{P.E.~Garrett$^{10}$}

\author{A.~Goldkuhle$^{11}$}

\author{C.~Henrich$^{12}$}

\author{A.~Illana$^8$}

\author{K.~Johnston$^2$}

\author{D.T.~Joss$^1$}

\author{J.M.~Keatings$^3$}

\author{N.A.~Kelly$^3$}

\author{M.~Komorowska$^{13}$}

\author{T.~Kr{\"o}ll$^{12}$}

\author{M.~Lozano$^2$}

\author{B.S.~Nara Singh$^3$}

\author{D.~O\textsc{\char13}Donnell$^3$}

\author{J.~Ojala$^{14,15}$}

\author{R.D.~Page$^1$}

\author{L.G.~Pedersen$^{16}$}

\author{C.~Raison$^{17}$}

\author{P.~Reiter$^{11}$}

\author{J.A.~Rodriguez$^2$}

\author{D.~Rosiak$^{11}$}

\author{S.~Rothe$^2$}

\author{M.~Seidlitz$^{11}$}

\author{T.M.~Shneidman$^{18}$}

\author{B.~Siebeck$^{11}$}

\author{J.~Sinclair$^3$}

\author{M.~Stryjczyk$^9$}

\author{P.~Van Duppen$^9$}

\author{S.~Vinals$^{19}$}

\author{V.~Virtanen$^{14,15}$}

\author{N.~Warr$^{11}$}

\author{K.~Wrzosek-Lipska$^{13}$}

\author{M. Zieli{\'n}ska$^{20}$}

\affiliation{
$^1$\mbox{University of Liverpool, Liverpool L69 7ZE, United Kingdom}\\
$^2$\mbox{ISOLDE, CERN,  1211 Geneva 23, Switzerland}\\
$^3$\mbox{University of the West of Scotland, Paisley PA1 2BE, United Kingdom}\\
$^4$\mbox{University of the Western Cape, Private Bag X17, Bellville 7535, South Africa}\\
$^5$\mbox{TRIUMF, Vancouver V6T 2A3 BC, Canada}\\
$^6$\mbox{Lund University, Box 118, Lund SE-221 00, Sweden}\\
$^7$\mbox{University of Michigan, Ann Arbor, Michigan 48109, USA}\\
$^8$\mbox{INFN Laboratori Nazionali di Legnaro, Legnaro 35020 PD, Italy}\\
$^9$\mbox{KU Leuven, Leuven B-3001, Belgium}\\
$^{10}$\mbox{University of Guelph, Guelph N1G 2W1 Ontario, Canada}\\
$^{11}$\mbox{University of Cologne, Cologne 50937, Germany}\\
$^{12}$\mbox{Technische Universit{\"a}t Darmstadt, Darmstadt 64289, Germany}\\
$^{13}$\mbox{Heavy Ion Laboratory, University of Warsaw, Warsaw PL-02-093, Poland}\\
$^{14}$\mbox{University of Jyvaskyla, P.O. Box 35, Jyvaskyla FIN-40014, Finland}\\
$^{15}$\mbox{Helsinki Institute of Physics, P.O. Box 64, Helsinki, FIN-00014, Finland}\\
$^{16}$\mbox{University of Oslo, P.O. Box 1048, Oslo N-0316, Norway}\\
$^{17}$\mbox{University of York, York YO10 5DD, United Kingdom}\\
$^{18}$\mbox{Joint Institute for Nuclear Research, RU-141980 Dubna, Russian Federation}\\
$^{19}$\mbox{Consejo Superior De Investigaciones Cient{\'i}ficas, Madrid S28040, Spain}\\
$^{20}$\mbox{IRFU CEA, Universit{\'e} Paris-Saclay, Gif-sur-Yvette F-91191, France}\\
}

\begin{abstract}
There is a large body of evidence that atomic nuclei can undergo
octupole distortion and assume the shape of a pear.  This phenomenon
is important for measurements of electric-dipole moments of atoms,
which would indicate CP violation and hence probe physics beyond the
standard model of particle physics. Isotopes of both radon and
radium have been identified as candidates for such measurements.
Here, we have observed the low-lying quantum states in $^{224}$Rn
and $^{226}$Rn by accelerating beams of these radioactive nuclei. We
report here additional states not assigned in our 2019 publication.
We show that radon isotopes undergo octupole vibrations but do not
possess static pear-shapes in their ground states. We conclude that
radon atoms provide less favourable conditions for the enhancement
of a measurable atomic electric-dipole moment.
\end{abstract}

\maketitle

\section{Introduction}

It is well established by the observation of rotational bands that
atomic nuclei can assume quadrupole deformation with axial and
reflection symmetry, usually with the shape of a rugby ball. The
distortion arises from long-range correlations between valence
nucleons which becomes favourable when the proton and/or neutron
shells are partially filled.  For certain values of proton and
neutron number it is expected that additional correlations will
cause the nucleus to also assume an octupole shape (`pear-shape')
where it loses reflection symmetry in the intrinsic
frame~\cite{butl96}. The fact that some nuclei can have pear shapes
has influenced the choice of atoms having nuclei with odd nucleon
number $A (= Z + N)$ employed to search for permanent
electric-dipole moments (EDMs). Any measurable moment will be
amplified if the nucleus has octupole collectivity and further
enhanced by static-octupole deformation. At present, experimental
limits on EDMs, that would indicate charge-parity (CP) violation in
fundamental processes where flavour is unchanged, have placed severe
constraints on many extensions of the Standard Model. Recently, new
candidate atomic species, such as radon and radium, have been
proposed for EDM searches. For certain isotopes octupole effects are
expected to enhance, by a factor 100-1000, the nuclear Schiff moment
(the electric-dipole distribution weighted by radius squared) that
induces the atomic EDM~\cite{spev97,doba05,elli11}, thus improving
the sensitivity of the measurement. There are two factors that
contribute to the greater electrical polarizability that causes the
enhancement: (i) the odd-A nucleus assumes an octupole shape; (ii)
an excited state lies close in energy to the ground state with the
same angular momentum and intrinsic structure but opposite parity.
Such parity doublets arise naturally if the deformation is static
(permanent octupole deformation).

The observation of low-lying quantum states in many nuclei with even
$Z, N$ having total angular momentum (`spin') and parity of
$I^\pi=3^-$ is indicative of their undergoing octupole vibrations
about a reflection-symmetric shape. Further evidence is provided by
the sizeable value of the electric octupole ($E$3) moment for the
transition to the ground state, indicating collective behaviour of
the nucleons. However the number of observed cases where the
correlations are strong enough to induce a static pear-shape is much
smaller.  Strong evidence for this type of deformation comes from
the observation of a particular behaviour of the energy levels for
the rotating quantum system and from an enhancement in the $E$3
moment~\cite{butl16}.  So far there are only two cases,
$^{224}$Ra~\cite{gaff13} and $^{226}$Ra~\cite{woll93} for which both
experimental signatures have been observed. The presence of a parity
doublet of 55 keV at the ground state of $^{225}$Ra makes this
nucleus therefore a good choice for EDM searches~\cite{bish16}. In
contrast to the radium isotopes, much less is known about the
behaviour of radon (Rn) nuclei proposed as candidates for atomic EDM
searches on account of possible enhancement of their Schiff
moments~\cite{dzub02,doba18,tard08,rand11,tard14,behr09,enge13,yama17,chup19}.
For this reason, different isotopes of radon have been listed in the
literature, e.g. $^{221,223,225}$Rn~\cite{behr09}, each having
comparable half-lives and ground state properties. The most commonly
chosen isotope for theoretical calculations~\cite{dzub02,doba18} and
the planning of experiments~\cite{tard08,rand11,tard14} is
$^{223}$Rn.

In our previous work~\cite{butl19} we presented new data on the
energy levels of heavy even-even Rn isotopes to determine whether
parity doublets are likely to exist near the ground state of
neighbouring odd-mass Rn nuclei. In this publication we report the
assignment of additional energy levels not presented in our previous
publication~\cite{butl19}. The direct observation of low-lying
states in odd-$A$ Rn nuclei (for example following Coulomb
excitation or $\beta$-decay from the astatine parent) presents
challenges, requiring advances in the technology used to produce
radioactive ions. We observe that $^{224,226}$Rn behave as octupole
vibrators in which the octupole phonon is aligned to the rotational
axis. We conclude that there are no isotopes of radon that have
static octupole deformation, so that any parity doublets in the
odd-mass neighbours will not be closely spaced in energy. This means
that radon atoms will provide less favourable conditions for the
enhancement of a measurable atomic electric-dipole moment.

\section{Results}

\subsection{Measurement of the quantum structure of heavy radon
isotopes}

In the experiments described here, $^{224}$Rn ($Z = 86, N = 138$)
and $^{226}$Rn ($Z = 86, N = 140$) ions were produced by spallation
in a thick thorium carbide target bombarded by $\approx 10^{13}$
protons s$^{-1}$ at 1.4 GeV from the CERN PS Booster. The ions were
accelerated in HIE-ISOLDE to an energy of 5.08 MeV per nucleon and
bombarded secondary targets of $^{120}$Sn.  In order to verify the
identification technique, another isotope of radon, $^{222}$Rn, was
accelerated to 4.23 MeV/u. The  $\gamma$-rays emitted following the
excitation of the target and projectile nuclei were detected in
Miniball~\cite{warr13}, an array of 24 high-purity germanium
detectors, each with six-fold segmentation and arranged in eight
triple-clusters. The scattered projectiles and target recoils were
detected in a highly segmented silicon detector~\cite{ostr02}. See
Methods.

Prior to our previous~\cite{butl19} and present work, nothing was
known about the energies and spins of excited states in
$^{224,226}$Rn, while de-exciting $\gamma$-rays from states in
$^{222}$Rn had been observed~\cite{sing11} with certainty up to
$I^\pi=13^-$.  The chosen bombarding energies for $^{224,226}$Rn
were about 3\% below the nominal Coulomb barrier energy at which the
beam and target nuclei come close enough in head-on collisions for
nuclear forces to significantly influence the reaction mechanism.
For such close collisions the population of high-spin states will be
enhanced, allowing the rotational behaviour of the nucleus to be
elucidated. This is the method first described by Ward et
al.~\cite{ward96} and has subsequently been coined unsafe Coulomb
excitation~\cite{wied99} as the interactions between the high-$Z$
reaction partners is predominantly electromagnetic. It is not
possible to precisely determine electromagnetic matrix elements
because of the small nuclear contribution. The most intense excited
states expected to be observed belong to the positive-parity
rotational band, built upon the ground state. These states are
connected by fast electric quadrupole ($E$2) transitions.  In nuclei
that are unstable to pear-shaped distortion, the other favoured
excitation paths are to members of the octupole band,
negative-parity states connected to the ground-state band by strong
$E$3 transitions.

\begin{figure}
\includegraphics[width=0.8\columnwidth]{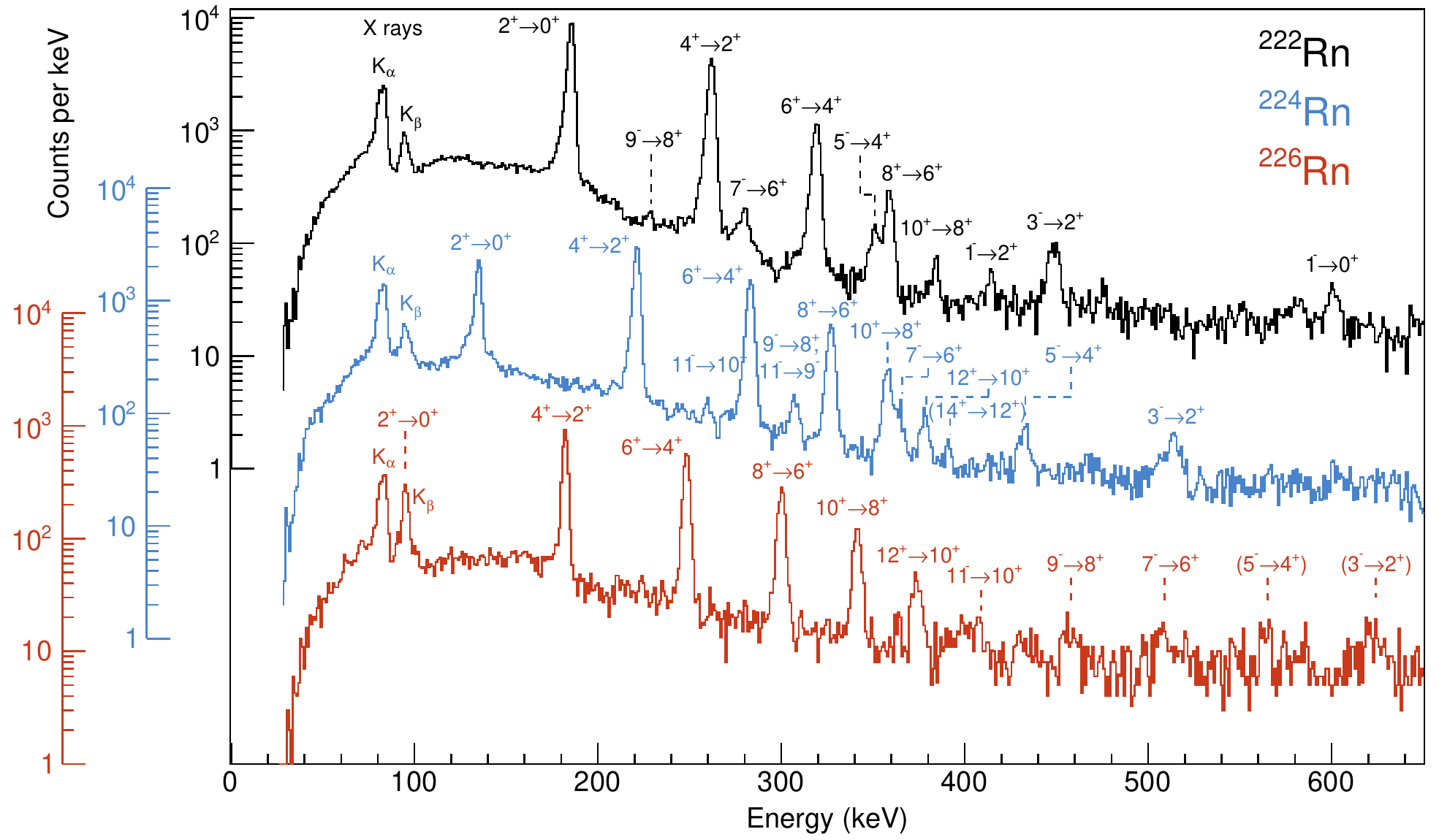}
\caption{\label{spectra}Spectra of  $\gamma$-rays. The $\gamma$-rays
were emitted following the bombardment of $^{120}$Sn targets by
$^{222}$Rn (black), $^{224}$Rn (blue), and $^{226}$Rn (red). The
$\gamma$-rays were corrected for Doppler shift assuming that they
are emitted from the scattered projectile. Random coincidences
between Miniball $\gamma$-ray and silicon particle detectors have
been subtracted. The transitions which give rise to the observed
full-energy peaks are labelled by the spin and parity of the initial
and final quantum states. The assignments of the transitions from
the negative parity states in $^{224,226}$Rn are tentative (see
text).}
\end{figure}

The spectra of  $\gamma$-rays time-correlated with scattered beam
and target recoils are shown in Fig.~\ref{spectra}.  The $E$2
$\gamma$-ray transitions within the ground-state positive-parity
band can be clearly identified, as these de-excite via a regular
sequence of strongly-excited states having spin and parity
$0^+,2^+,4^+,...$ with energies $\frac{\hbar^2}{2\cal{J}} I(I+1)$.
In this expression the moment-of-inertia $\cal{J}$ systematically
increases with increasing $I$ (reducing pairing) and with number of
valence nucleons (increasing quadrupole deformation). As expected
from multi-step Coulomb excitation the intensities of the
transitions systematically decrease with increasing $I$, after
correcting for internal conversion and the  $\gamma$-ray detection
efficiency of the Miniball array.

\begin{figure}
\includegraphics[width=0.8\columnwidth]{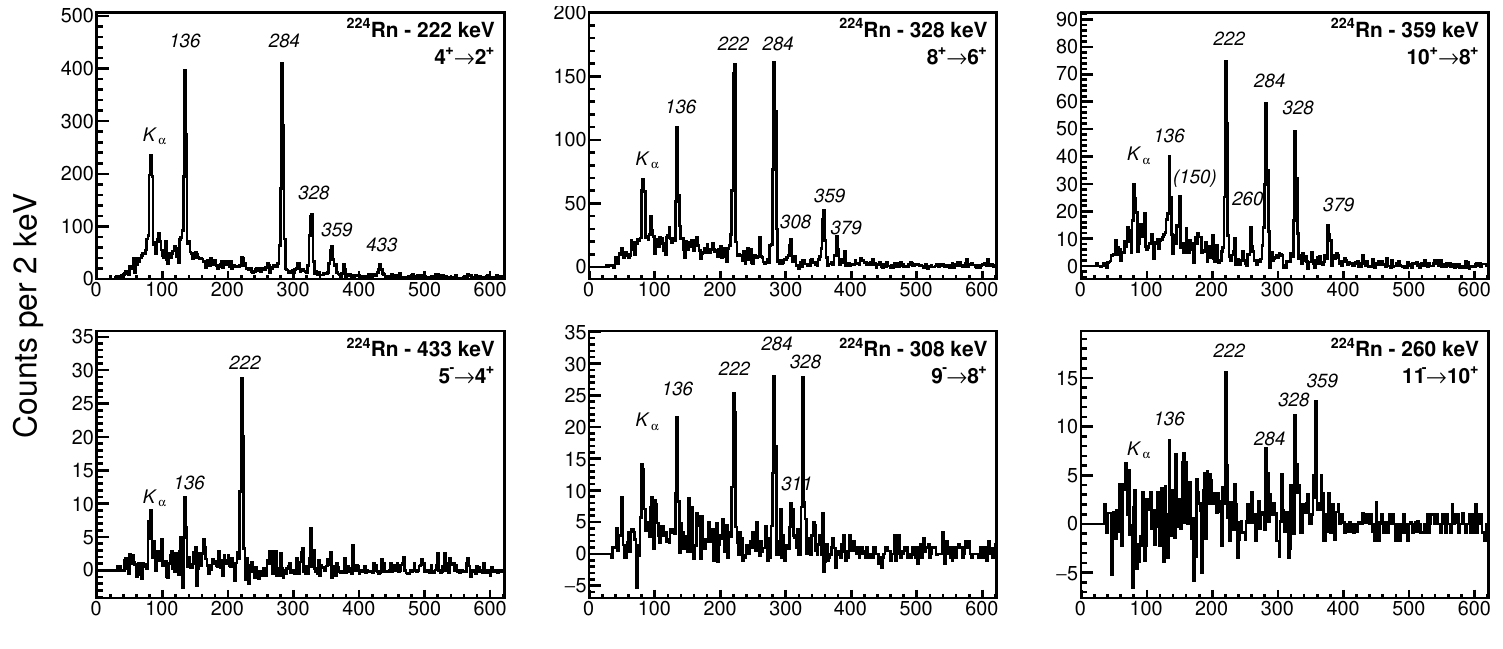}
\includegraphics[width=0.8\columnwidth]{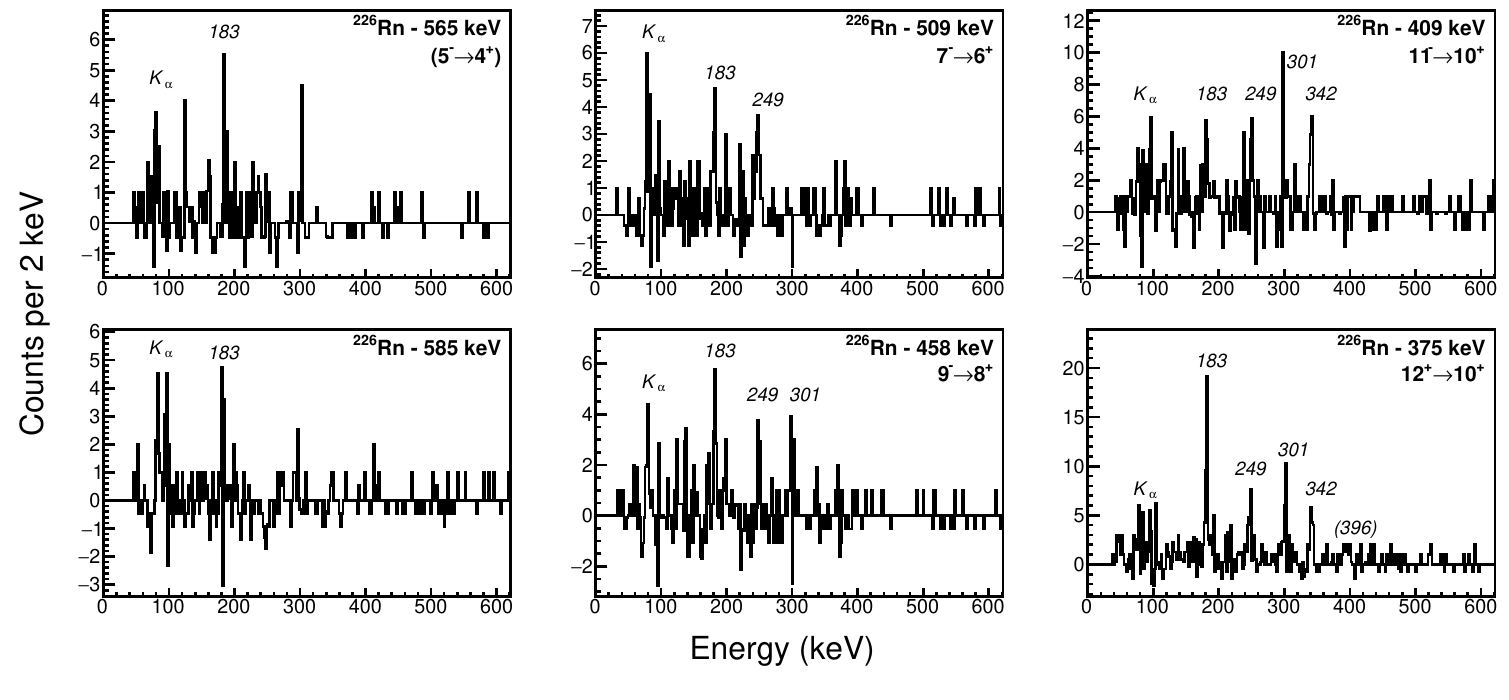}
\caption{\label{gates}Coincidence  $\gamma$-ray spectra.  The
representative background-subtracted  $\gamma$-ray spectra are in
time-coincidence with different gating transitions.  Here the
observed peaks are labelled by the energy (in keV) of the
transition. The gating transition is additionally labelled by the
proposed spin and parity of the initial and final states.}
\end{figure}

The other relatively intense $\gamma$-rays observed in these spectra
with energies $< 600$ keV are assumed to have electric-dipole ($E$1)
multipolarity, and to depopulate the odd-spin negative-parity
members of the octupole band. In order to determine which states are
connected by these transitions, pairs of time-correlated
(`coincident')  $\gamma$-rays were examined. In this analysis, the
energy spectrum of  $\gamma$-rays coincident with one particular
transition is generated by requiring that the energy of this
`gating' transition lies in a specific range. Typical spectra
obtained this way are shown in Fig.~\ref{gates}. Each spectrum
corresponds to a particular gating transition, background
subtracted, so that the peaks observed in the spectrum arise from
$\gamma$-ray transitions in coincidence with that transition.

\begin{figure}
\centering
\includegraphics[width=0.9\columnwidth]{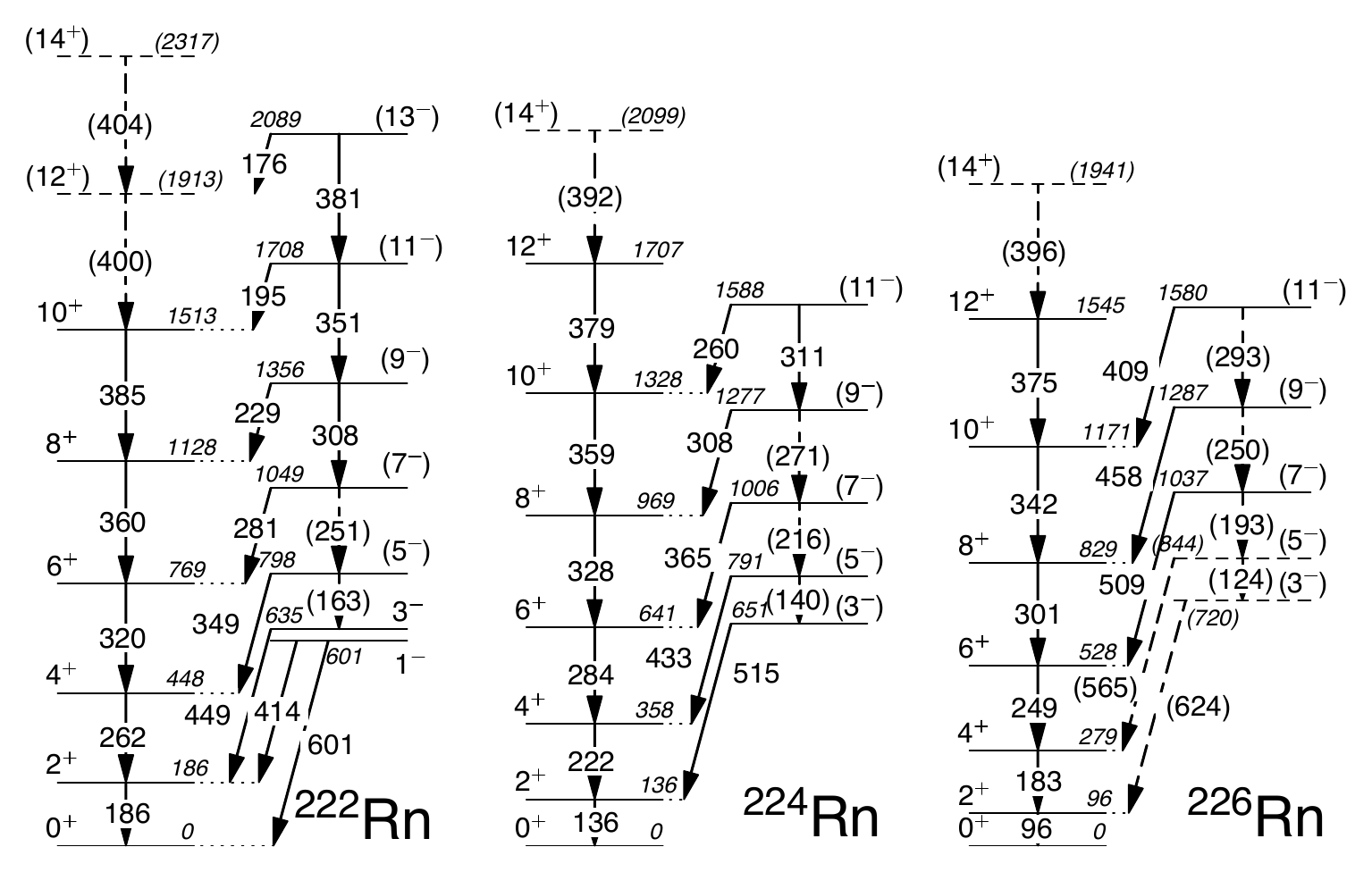}
\caption{\label{level schemes}Level schemes. These partial
level-schemes for $^{222,224,226}$Rn show the excited states of
interest. Arrows indicate  $\gamma$-ray transitions. All energies
are in keV. Firm assignments of transitions and energy levels in the
scheme are from previous work~\cite{sing11} or have been made using
$\gamma$-$\gamma$-coincidence relations; tentatively assigned
transitions such as those from the $3^-$ and $5^-$ states in
$^{226}$Rn are shown as dashed lines. The assignment of spins and
parities in brackets are tentative.}
\end{figure}

These $\gamma$-$\gamma$ spectra contain significantly more data than
those reported previously~\cite{butl19}, see Methods. The additional
$\gamma$-$\gamma$ data have allowed us to extend the level schemes
by one additional state in each of the positive-parity bands in
$^{224,226}$Rn and in the negative-parity band in $^{224}$Rn
compared to that published previously~\cite{butl19}. More
importantly, we are able to determine the probable energy of the
$7^-$, $9^-$ and $11^-$ states in $^{226}$Rn, see Fig.~\ref{gates}.
By extrapolating this band to lower spin states on the basis of its
rotational-like behaviour, we are able to estimate the energies of
the $5^-$ and $3^-$ states, whose decays to the positive-parity band
are observed in the total $\gamma$-ray spectrum, see
Fig.~\ref{spectra}. These tentative assignments imply that the
energy of the $5^- \rightarrow 4^+$ transition is 565~keV, not
585~keV as reported previously~\cite{butl19}. The gated coincidence
spectra for both candidate $5^- \rightarrow 4^+$ transitions are
shown in Fig.~\ref{gates}. In both cases the quality of the gated
spectra does not allow a firm assignment to be made to the
$\gamma$-ray transition.  The spectrum gated by the candidate $3^-
\rightarrow 2^+$ transition is featureless, as expected given the
large internal conversion of the $2^+ \rightarrow 0^+$ transition.
The level schemes for $^{224,226}$Rn constructed from the
coincidence spectra, together with the known~\cite{sing11} scheme
for $^{222}$Rn, are shown in Fig.~\ref{level schemes}.  For
$^{226}$Rn the energy of the strongly-converted $2^+ \rightarrow
0^+$ transition overlaps with those of the $K_\beta$ X-rays, but its
value can be determined assuming that the relative intensity of
$K_\beta$, $K_\alpha$ X-rays is the same as for $^{222,224}$Rn. The
$E$2 transitions connecting the states in the octupole band are not
observed because they cannot compete with faster, higher-energy $E$1
decays.  The only other plausible description for this band is that
it has $K^\pi=0^+$ or $2^+$, implying that the $K^\pi=0^-$ octupole
band is not observed. (Here $K$ is the projection of
\boldmath$I$\unboldmath ~on the body-fixed symmetry axis.) This is
unlikely as the bandhead would have to lie significantly lower in
energy than has been observed in $^{222,224,226}$Ra, and inter-band
transitions from states with $I'
> 4$ to states with $I$ and $I-2$ in the ground-state band and in-band
transitions to $I'-2$ would all be visible in the spectra. The spin
and parity assignments for the positive-parity band that is strongly
populated by Coulomb excitation can be regarded as firm, whereas the
negative-parity state assignments are made in accord with the
systematic behaviour of nuclei in this mass region.

\subsection{Characterisation of octupole instability}

\begin{figure}
\includegraphics[width=0.8\columnwidth]{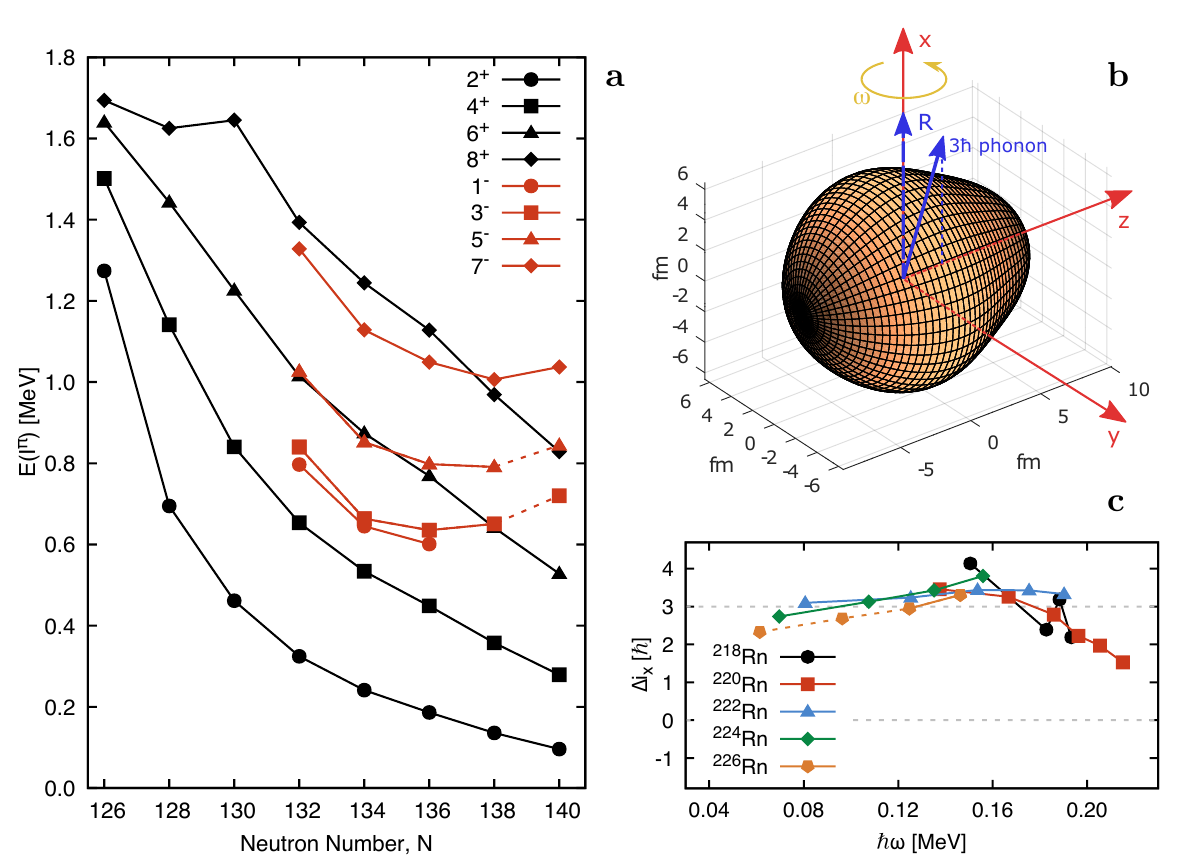}
\caption{\label{systematics}Systematic behaviour of radon isotopes.
{\bf a} Systematics of the energies for different spins of low-lying
positive-parity (black) and negative-parity states (red) in radon
isotopes; {\bf b}  cartoon illustrating how the octupole phonon vector
aligns with the rotation ($R$) vector (which is orthogonal to the
rotating body's symmetry axis) so that $I=R+3 \hbar$ and $\Delta i_x
= 3 \hbar$; {\bf c} difference in aligned spin for negative- and
positive-parity states in $^{218-226}$Rn (re-analysed for
$^{218-222}$Rn that have been presented earlier\cite{cock99}). The
dashed line at $\Delta i_x=0$ is the expected value for static
octupole deformation. }
\end{figure}

From the level schemes and from the systematics for all the radon
isotopes (Fig.~\ref{systematics}) it is clear that the bandhead of
the octupole band reaches a minimum around $N=136$.  The character
of the octupole bands can be explored~\cite{cock99} by examining the
difference in aligned angular momentum, $\Delta i_x = i_x^- -
i_x^+$, at the same rotational frequency $\omega$, as a function of
$\omega$. Here $i_x$ is approximately $I$ for $K=0$ bands and $\hbar
\omega$ is approximately $(E_I - E_{(I-2)})/2$.   For nuclei with
permanent octupole deformation $\Delta i_x$ is expected to approach
zero, as observed for several isotopes of Ra, Th, and
U~\cite{butl16}. For octupole vibrational nuclei in which the
negative-parity states arise from coupling an octupole phonon to the
positive-parity states, it is expected that $\Delta i_x \approx 3
\hbar$ as the phonon prefers to align with the rotational axis. This
is the case for the isotopes $^{218,220,222,224,226}$Rn at values of
$\hbar \omega$ ($< 0.2$ MeV) where particle-hole excitations do not
play a role, see Fig.~\ref{systematics}. Thus we have clearly
delineated the lower boundary at $Z > 86$ as to where permanent
octupole deformation occurs in nature.

\section{Discussion}

The observation of octupole-vibrational bands in the even-even radon
isotopes is consistent with several theoretical
calculations~\cite{naza84,robl13,agbe16}, which predict that only
nuclei with $Z > 86$ have stable octupole deformation. Other
calculations suggest that radon isotopes with $A \approx 222$ will
have non-zero values of the octupole deformation parameter
$\beta_3$~\cite{moll08,robl12}. For such nuclei, which have a
minimum in the nuclear potential energy at non-zero values of
$\beta_3$, the positive- and negative-parity states are projected
from intrinsic configurations having  $K^\pi = 0^+,0^-$ which are
degenerate in energy. In the odd-$A$ neighbours parity doublets
arise by coupling the odd particle to these configurations. This is
not the case for reflection-symmetric nuclei that undergo octupole
vibrations around $\beta_3 = 0$.  Bands of opposite parity with
differing single-particle configurations can lie close to each other
fortuitously~\cite{verm90,wisn17} but in general those arising from
coupling the odd nucleon to the ground state and octupole phonon
will be well separated.  The separation will be determined by the
spacing of the bands in the even-even core, $\approx 500$ keV in the
case of $^{222-226}$Rn (see Fig.~\ref{systematics}), and will be in
general much larger than that the value ($\approx 50$ keV) observed
for parity doublets in radium isotopes~\cite{butl96}. Realistic
estimates of Schiff moments for octupole-vibrational systems have
yet to be made~\cite{auer06,zele08}. Nevertheless, it can be
concluded that, if measurable CP-violating effects occur in nuclei,
the enhancement of nuclear Schiff moments arising from octupole
effects in odd-$A$ radon nuclei is likely to be much smaller than
for heavier octupole-deformed systems.

\section{METHODS}

\subsection{Production of radioactive radon beams}

In our experiments, $^{222,224,226}$Rn were produced by spallation
in the primary target, diffused to the surface and then singly
ionized ($q = 1^+$) in an enhanced plasma ion-source~\cite{pene10}
with a cooled transfer line. The ions were then accelerated to 30
keV, separated according to $A/q$, and delivered to a Penning trap,
REXTRAP~\cite{wolf03}, at a rate of around $8 \times 10^6$ ions
s$^{-1}$ for $^{222}$Rn, $2 \times 10^6$ ions s$^{-1}$ for
$^{224}$Rn and $10^5$ ions s$^{-1}$ for $^{226}$Rn at the entrance.
Inside the trap, the singly-charged ions were accumulated and cooled
before being allowed to escape in bunches at 500 ms intervals into
an electron-beam ion source, REXEBIS~\cite{wolf03}. Here, the ions
were confined for 500-700 ms in a high-density electron beam that
stripped more electrons to produce a charge state of $51^+$
($^{222}$Rn) or $52^+$ ($^{224,226}$Rn) extracted as 1 ms pulses
before being mass-selected again according to $A/q$, and injected at
2 Hz into the HIE-ISOLDE linear post-accelerator. The Rn beams then
bombarded a $^{120}$Sn target of thickness $\approx 2$ mg cm$^{-2}$
with an intensity of about $6 \times 10^5$ ions s$^{-1}$, $1.1
\times 10^5$ ions s$^{-1}$ and $2 \times 10^3$ ions s$^{-1}$ for
$^{222}$Rn, $^{224}$Rn and $^{226}$Rn, respectively. The total
beam-times were respectively 8, 16 and 24h. The level of Fr impurity
in the Rn beams could be estimated for $A=222$ as below 1\% by
observing radioactive decays at the end of the beam line.

\subsection{Data selection and Doppler correction}

Events corresponding to the simultaneous detection of  $\gamma$-rays
and heavy ions in Miniball and the silicon detector array
respectively were selected if the measured energy and angle of
either projectile or target satisfied the expected kinematic
relationship for inelastic scattering reactions. This procedure
eliminated any background from stable noble-gas contaminant beams
produced in REX-TRAP having the same $A/q$ as the radon beams. In
the present setup the average angle of each of the 16 strips of one
side of the silicon detector array ranged between $19.6^\circ$ and
$54.9^\circ$ to the beam direction, corresponding to a scattering
angular range of $140.9^\circ$ to $70.2^\circ$ in the
centre-of-mass. In order to reduce background from Compton
scattering, events were rejected if any two germanium crystals in
each triple-cluster registered simultaneous $\gamma$-ray hits, in
contrast to the normal adding procedure which would substantially
increase the probability of summing two $\gamma$-rays emitted in the
same decay sequence  (`true pile-up'). Miniball was calibrated using
$^{133}$Ba and $^{152}$Eu radioactive sources that emitted
$\gamma$-rays of known energy and relative intensity. The
relativistic Doppler correction was performed by determining the
momentum vector of the projectile, using the energy and position
information in the pixelated silicon detector, and the emission
polar and azimuthal angle of the detected $\gamma$-ray in the
segment of Miniball where most energy was deposited. In the case of
the latter the relative orientation of each segment to each other
and to the beam axis was determined by employing
d($^{22}$Ne,p$\gamma$) and d($^{22}$Ne,n$\gamma$) reactions. The
Doppler-corrected energies for transitions in $^{224}$Rn and
$^{226}$Rn together with the deduced level energies are given in
Table~\ref{energies}. Subsequent to our earlier
publication~\cite{butl19}, the sort code that converts raw data from
the Miniball spectrometer to Root analysis files
(MiniballCoulexSort, see Data availability, below) has been modified
so that $\gamma$-$\gamma$ data are included when both heavy ions are
detected in the silicon detector array, in addition to single-ion
events. Since both conditions for ion detection were already
considered for single $\gamma$-ray events, this modification does
not affect the total spectra shown in Fig.~\ref{spectra}, but
considerably enhances the statistics of the $\gamma$-$\gamma$ gated
spectra shown in Fig.~\ref{gates}.

\begin{table*}
\caption{\label{energies} Energies of levels and transitions in
$^{224}$Rn and $^{226}$Rn. The 1-$\sigma$  errors are given,
estimated from the
 statistical error and the uncertainty in the energy calibration and Doppler correction.}
\begin{ruledtabular}
\begin{tabular}{cccccccc}
$^{224}$Rn & & & & $^{226}$Rn & & &\\
$E_{level}$ & $I^\pi_i$ & $E_\gamma$ (keV) & $I^\pi_f$ & $E_{level}$ & $I^\pi_i$ & $E_\gamma$ (keV) & $I^\pi_f$ \\
\hline
135.6 (5) & $2^+$ & 135.6 (5)& $0^+$ & 96.0 (11) & $2^+$ & 96.0 (11) & $0^+$\\
357.6 (6) &  $4^+$  & 222.0 (5) &  $2^+$ & 278.9 (12) & $4^+$ & 182.9 (5) & $2^+$\\
641.4 (8) &  $6^+$  & 283.8 (5) &  $4^+$ &  527.9 (13) & $6^+$ & 249.0 (5) & $4^+$\\
969.2 (9) &  $8^+$ & 327.8 (5) &  $6^+$ & 828.6 (14) &  $8^+$ & 300.7 (5) & $6^+$\\
1327.8 (10) & $10^+$ & 358.6 (5) &  $8^+$  & 1170.8 (14) & $10^+$ & 342.1 (5) & $8^+$\\
1706.8 (11) & $12^+$ &   379.1 (5) & $10^+$ & 1545.4 (15) & $12^+$ & 374.6 (5) & $10^+$\\
(2098.7) (13) & ($14^+$) & (391.8) (6) & $12^+$ & (1941.2) (19) & ($14^+$)& (395.8) (11) & $12^+$\\
650.6 (8) &  ($3^-$) &   515.0 (6) &  $2^+$ &  (719.9) (17) & ($3^-$) & (623.9) (13) & $2^+$ \\
790.8 (8) &   ($5^-$) &   433.2 (5) & $4^+$ & (843.6) (14) & ($5^-$) & (564.7) (8) & $4^+$\\
1006.4 (10) & ($7^-$) &  365.0 (5) &  $6^+$ & 1036.8 (16) & ($7^-$) & 508.9 (9) & $6^+$\\
1277.2 (10) & ($9^-$) &   308.0 (5) & $8^+$ &  1286.7 (17) & ($9^-$) & 458.1 (10) & $8^+$\\
1588.3 (13) & ($11^-$) & 260.5 (8) & $10^+$ & 1579.6 (18) & ($11^-$) & 408.8 (11) & $10^+$\\

\end{tabular}
\end{ruledtabular}
\end{table*}

\begin{acknowledgments}
We are grateful to Niels Bidault, Eleftherios Fadakis, Erwin
Siesling, and Fredrick Wenander who assisted with the preparation of
the radioactive beams, and we thank Jacek Dobaczewski for useful
discussions.  The support of the ISOLDE Collaboration and technical
teams is acknowledged. We are grateful to Ralph Kern who initiated
the improvement to the Miniball sort code. This work was supported
by the following Research Councils and Grants: Science and
Technology Facilities Council (STFC; UK) grants ST/P004598/1,
ST/L005808/1; Federal Ministry of Education and Research (BMBF;
Germany) grants 05P18RDCIA, 05P15PKCIA and 05P18PKCIA and the
``Verbundprojekt 05P201''; National Science Centre (Poland) grant
2015/18/M/ST2/00523; European Union's Horizon 2020 Framework
research and innovation programme 654002 (ENSAR2); Marie
Sk{\l}odowska-Curie COFUND grant (EU-CERN) 665779; Research
Foundation Flanders (FWO, Belgium), by GOA/2015/010 (BOF KU Leuven)
and the Interuniversity Attraction Poles Programme initiated by the
Belgian Science Policy Office (BriX network P7/12); RFBR (Russia)
grant 17-52-12015; the Academy of Finland (Finland) Grant No.
307685.
\end{acknowledgments}

\section{Data availability}

The data that support the findings of this study are available from
the corresponding author on reasonable request. The software used to
sort the raw data is available at doi:10.5281/zenodo.2593370
(Gaffney, L. P. and Konki, J., MiniballCoulexSort for Coulex, SPEDE,
CREX and TREX). Information about the ROOT software package used to
analyse the data can be found at https://root.cern.ch/publications .

\end{document}